\documentclass[12pt,twoside]{article}

\evensidemargin=\oddsidemargin
\def\Title#1#2#3{%
    \baselineskip=18pt
    \begin{center}
          {\large\bf{#1} \\ }
          \bigskip\bigskip
          {#2} \\
          {#3} \\
    \end{center}}
\long\def\Abstract#1{%
         \bigskip
         \parbox{0.93\textwidth}{%
                 \begin{center}
                       {\bf Abstract} \\
                 \end{center}
                 \medskip{\baselineskip=14pt #1}
                 \vss}
         \bigskip}

\newcommand{\myfigure}[1]{%
         \framebox[50mm]{\epsfxsize=48mm\epsfbox{#1}}}
\input epsf

\makeatletter
\renewcommand{\section}%
 {\@startsection{section}{1}{0pt}%
  {-3.25ex plus -1ex minus -.2ex}{1.5ex plus .2ex}%
  {\vspace*{5mm}\raggedright\large\bf }}
\renewcommand{\thesection}{\arabic{section}.}
\renewcommand{\thefigure}{\arabic{figure}.}
\@addtoreset{equation}{section}
\renewcommand{\@eqnnum}{(\thesection\theequation)}
\renewcommand{\p@equation}{\thesection}
\makeatother

\begin{document}

\Title{Quantum cosmological solutions:\\
their dependence on the choice of gauge conditions\\
and physical interpretation}%
{T. P. Shestakova}%
{Department of Theoretical and Computational Physics,
Southern Federal University\footnote{former Rostov State University},\\
Sorge St. 5, Rostov-on-Don 344090, Russia \\
E-mail: {\tt shestakova@phys.rsu.ru}}

\Abstract{In ``extended phase space'' approach to quantum
geometrodynamics numerical solutions to Schr\"odinger equation
corresponding to various choice of gauge conditions are obtained
for the simplest isotropic model. The ``extended phase space''
approach belongs to those appeared in the last decade in which, as
a result of fixing a reference frame, the Wheeler -- DeWitt static
picture of the world is replaced by evolutionary quantum
geometrodynamics. Some aspects of this approach were discussed at
two previous PIRT meetings. We are interested in the part of the
wave function depending on physical degrees of freedom. Three gauge
conditions having a clear physical meaning are considered. They are
the conformal time gauge, the gauge producing the appearance of
$\Lambda$-term in the Einstein equations, and the one covering the
two previous cases as asymptotic limits. The interpretation and
discussion of the obtained solutions is given.}

\section{Introduction}
In this paper we present solutions to quantum geometrodynamical
Schr\"odinger equation corresponding to various choice of gauge
conditions for the simplest isotropic model. It is widely accepted
in quantum geometrodynamics to illustrate general ideas taking
simple cosmological models as examples. The reason why physicists
working in this field appeal to simple models is that now quantum
geometrodynamics is just as far from being a completed theory as it
was decades ago. One must confess that hitherto there is no
agreement on what "first principles" this theory should be based
and what is the form of master equation for a wave function of the
Universe. The first version of quantum geometrodynamics, proposed
by Wheeler and DeWitt \cite{DeWitt,Wheeler}, encountered a number
of fundamental problems (for discussion, see \cite{Isham,SS1,
SS2}). The main problem is the so called ``frozen formalism'', or
the absence of time evolution. It is easy to see that the source of
the problem of time consists in the application of the Dirac
postulates to gravitational field, according to which not the
Schr\"odinger equation but the constraints as conditions on a wave
function play the central part in the theory. As a result of
impossibility to resolve the problems of the Wheeler -- DeWitt
quantum geometrodynamics in its own limits, in the last decade
there appear a new tendency in the development of the theory which
can be called Evolutionary Quantum Gravity. This tendency may be
characterized by the two features: firstly, the recognition of the
fact that it is impossible to obtain the evolutionary picture of
the Universe without fixing a reference frame and, secondly, the
rejection of the Wheeler -- DeWitt equation and the reestablishment
of the role which the Schr\"odinger equation plays in any quantum
theory.

The tendency embraces several approaches (see, for example,
\cite{BK,BM}, where a dust fluid is considered as a good choice to
fix a reference frame in quantum gravity), to which the ``extended
phase space'' approach belongs. Some aspects of the latter were
discussed at two previous PIRT meetings \cite{Shest1,Shest2}. The
approach is based on a careful analysis of peculiarities of
quantization of the Universe as a whole \cite{SSV1,SSV2}. The
analysis showed that quantum geometrodynamics as a mathematically
consistent theory failed to be constructed in a gauge invariant
way, therefore, the Wheeler -- DeWitt equation, being a constraint
on a state vector, loses its significance and should be replaced by
a gauge dependent Schr\"odinger equation resulting from the
Hamiltonian formulation of the theory in extended phase space. A
wave function satisfying the Schr\"odinger equation is determined
on extended configurational space that involves gauge gravitational
degrees of freedom equally as physical ones. However, we are
actually interested in the part of the wave function depending on
physical degrees of freedom only, since this very function defines
probability distributions of physical quantities.

In Section 2 we shall describe the model and the Schr\"odinger
equation for the physical part of wave function for the given
model. Since the form of the Schr\"odinger equation is gauge
dependent, to obtain descriptions of the Universe corresponding to
various gauge conditions (in other words, to various reference
frames) one has to solve, in fact, absolutely different
differential equations. It naturally leads us to the question, is
there any correspondence among solutions of the equations? And how
should they be interpreted?

Let us note that while in \cite{BK,BM} the authors work with a
certain parametrization of gravitational variables (as a rule, it
is the Arnowitt -- Deser -- Misner parametrization \cite{ADM}) and
some ``privileged'' reference frame, our approach, though was
applied to cosmological models with finite degrees of freedom,
aimed at including arbitrary parametrizations and a wide enough
class of gauge conditions. We shall consider three gauge conditions
having a clear physical meaning: the conformal time gauge, the
gauge producing the appearance of $\Lambda$-term in the Einstein
equations, and the one covering the two previous cases as
asymptotic limits. For a closed universe, the first and third
gauges gives rise to a discrete Hamiltonian spectrum, while the
second gauge leads to a continuous spectrum. From a pure methodical
viewpoint, the first and third cases are much easier to be treated,
and in Section 3 numerical solution for these cases will be
presented, meantime the second case admits qualitative
consideration only. Section 4 contains physical interpretation and
conclusions.

\section{The model and the Schr\"odinger equation for the physical part
of the wave function}
The action for a closed isotropic universe is
\begin{equation}
\label{action-1}
S=-\!\int\!dt\,\left(\frac12\frac{a\dot a^2}N
  -\frac12Na\right)+S_{(mat)}+S_{(gf)},
\end{equation}
\begin{equation}
\label{action-mat-gf}
S_{(mat)}=-\!\int\!dt\,Na^3\varepsilon(a),\quad
S_{(gf)}=\!\int\!dt\,\pi_0\left(\dot N
 -\frac{d f}{d a}\dot a\right).
\end{equation}
Matter fields are described in this model phenomenologically,
without a clear indication on the nature of the fields. The
dependence of its energy density
$\varepsilon(a)$ on the scale factor $a$ determines
its equation of state, namely, for the power dependence
$\varepsilon(a)=\displaystyle\frac{\varepsilon_0}{a^n}$,
the equation of state is known to be
$p_{(mat)}=\left(\displaystyle\frac n3-1\right)\varepsilon_{(mat)}$,
$\varepsilon_0$ is a constant whose dimensionality in the Plank
units is $\rho_{Pl}l_{Pl}^n$. Since we are interested in
early enough stages of the Universe evolution, we shall suppose
that the Universe was filled with radiation with the equation of
state $p_{(mat)}=\displaystyle\frac13\varepsilon_{(mat)}$, i.e.
\begin{equation}
\label{eps-rad}
\varepsilon(a)=\frac{\varepsilon_0}{a^4}.
\end{equation}
$S_{(gf)}$ is a gauge-fixing part of the action, its variation
giving rise to gauge dependent terms in the Einstein equations. In
ordinary quantum theory this terms are to be excluded by asymptotic
boundary conditions. As was argued in \cite{SSV1}, in the case of
the Universe with a non-trivial topology, which, in general, does
not possess asymptotic states, making use of asymptotic boundary
condition is not justified.

If so, the gauge-fixing action describes a subsystem of the
Universe, some medium, whose state is determined by a chosen gauge.
In (\ref{action-mat-gf}) a differential form of the gauge condition
\begin{equation}
\label{gauge}
N-f(a)=0
\end{equation}
is used. The equation of state for this subsystem is
\begin{equation}
\label{eq-states-gen}
p_{(obs)}=\frac13\frac{f'(a)}{f(a)}a\varepsilon_{(obs)}.
\end{equation}
The index ({\it obs}) indicates that this subsystem corresponds to an
observer studying the Universe evolution in his reference frame.

The action (\ref{action-1}) is a particular case of the action for
a cosmological model with a finite number degrees of freedom
considered in \cite{Shest1,SSV2}. The Schr\"odinger equation for
the physical part of the wave function looks like
\begin{equation}
\label{Schr-eq-gen}
\left.\left[-\frac12\sqrt{\frac N a}\frac d{d a}
  \left(\sqrt{\frac N a}\frac{d\Psi}{d a}\right)
 +\frac12 N a\Psi-N a^3\varepsilon(a)\Psi\right]\right|_{N=f(a)}
 =E\Psi.
\end{equation}
From the classical point of view, $E$ is given by
\begin{equation}
\label{E-cons}
E=-\!\int\!\sqrt{-g}\,T_{0(obs)}^0\,d^3x.
\end{equation}
$T_{\mu(obs)}^{\nu}$ is a quasi energy-momentum tensor obtained by
variation of the gauge-fixing action; it is not a real tensor in
the sense that it depends on a gauge condition.
$T_{\mu(obs)}^{\nu}$ describes the subsystem of the observer in the
gauged Einstein equations \cite{Shest1}. It can be shown that the
integral (\ref{E-cons}) of $T_{0(obs)}^0$ taken over space is a
conserved quantity for the class of gauge conditions (\ref{gauge}).
Thus, $E$ characterizes the energy of the observer subsystem.

It may be said that on a phenomenological level this approach takes
into account interaction between the observer subsystem and the
physical Universe. The interaction causes rebuilding of energy
balance of two subsystems. It is expected that at the late stage of
the Universe evolution, when the Universe is well described by
General Relativity, gauge effects are negligible, and the values of
$E$ must be very close, if not equal, to zero. However, at the
early quantum stage $E$ may have essentially non-zero values, and
the exploration of its spectrum is the main task of this work.

Now we consider several gauge conditions.

1. The conformal time gauge $N=a$. The equation of state of the
observer subsystem is the same as that of the matter:
$p_{(obs)}=\displaystyle\frac13\varepsilon_{(obs)}$. Substituting
$N=a$ and (\ref{eps-rad}) in (\ref{Schr-eq-gen}), we get
\begin{equation}
\label{Schr-eq-1-1}
-\frac12\frac{d^2\Psi}{da^2}+\frac12a^2\Psi-\varepsilon_0\Psi=E\Psi.
\end{equation}
After redefinition
\begin{equation}
\label{E-redef}
E+\varepsilon_0\to E
\end{equation}
we obtain the equation
\begin{equation}
\label{Schr-eq-1}
-\frac12\frac{d^2\Psi}{da^2}+\frac12a^2\Psi=E\Psi.
\end{equation}
Therefore, Eq. (\ref{Schr-eq-1}) describes the Universe filled with
a ``substance'' with the equation of state
$p=\displaystyle\frac13\varepsilon$. Just some part of the energy
of this substance may be due to a usual matter while the other part
may be due to gauge, or observer, effects.

It was shown in \cite{Shest3} that Eq. (\ref{Schr-eq-1}) can be
obtained in the limits of the Wheeler -- DeWitt quantum
geometrodynamics by rewriting of the Wheeler -- DeWitt equation
$H\Psi=0$ as a Schr\"odinger-like equation $\tilde H\Psi=E\Psi$.
Under additional requirements, that imply choosing a certain gauge
condition and including a certain kind of matter into the model,
the classical Hamiltonian constraint $H=0$ can be presented in a
new form, $\tilde H=E$, $H=\tilde H-E$, where $E$ is a conserved
quantity which appears from phenomenological consideration of this
kind of matter. So, in this approach, $E=\varepsilon_0$, i.e. $E$
is entirely due to the usual matter (radiation).

On the other side, the need for making a choice of gauge to rewrite
the Wheeler -- DeWitt equation in the special form
$\tilde H\Psi=E\Psi$ witnesses to gauge noninvariance of the
Wheeler -- DeWitt theory. As was already emphasized above, the
Wheeler -- DeWitt equation loses its meaning, and it seems to be
reasonable rejecting it rather trying to hold it by any means.

The effective potential $U(a)=\displaystyle\frac12a^2$ is given at
Fig. 1(a).
\begin{center}
\tabcolsep=1mm
\begin{tabular}{|c|c|c|}
\hline
\rule{0mm}{50mm}
\myfigure{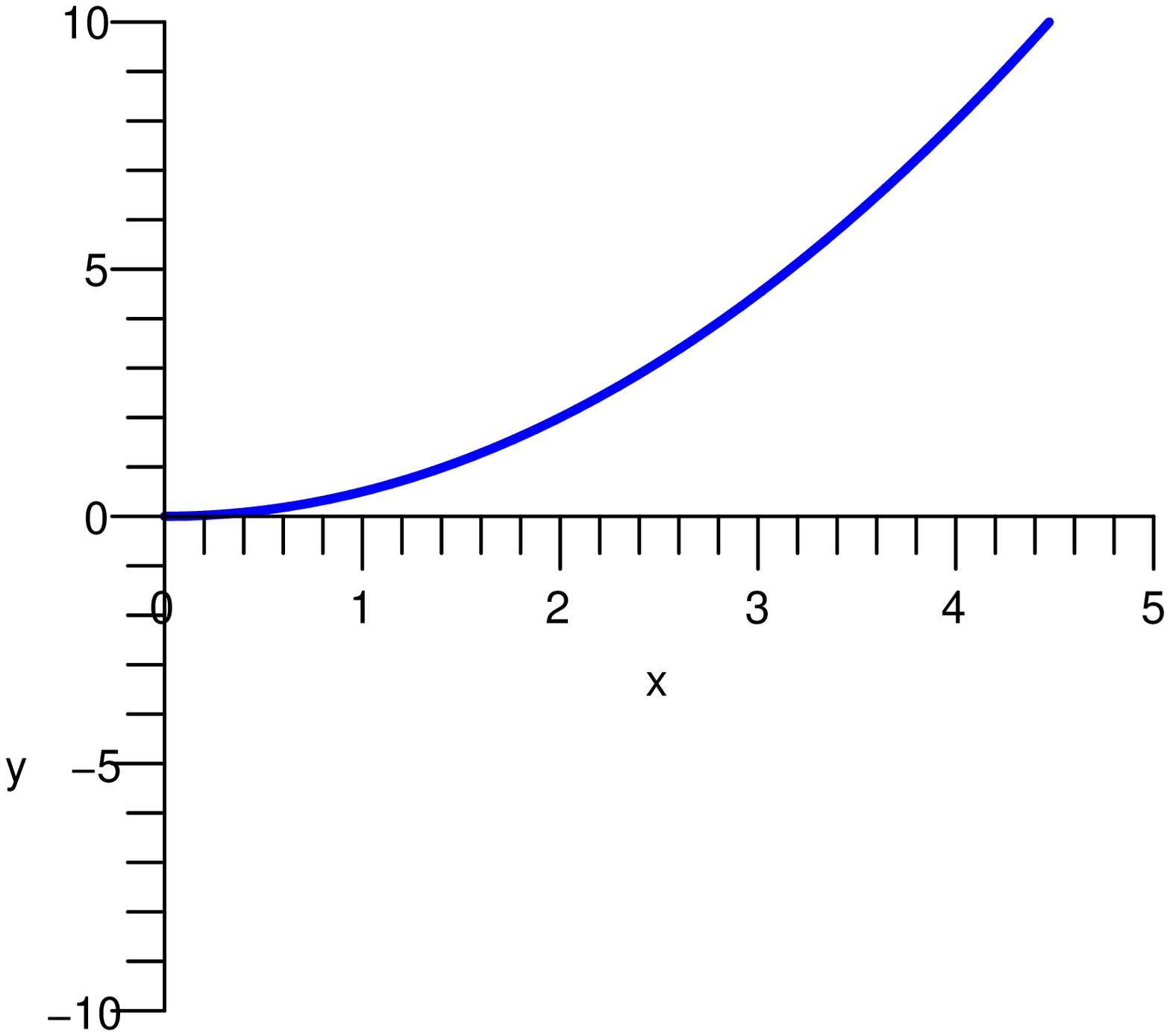}&
\myfigure{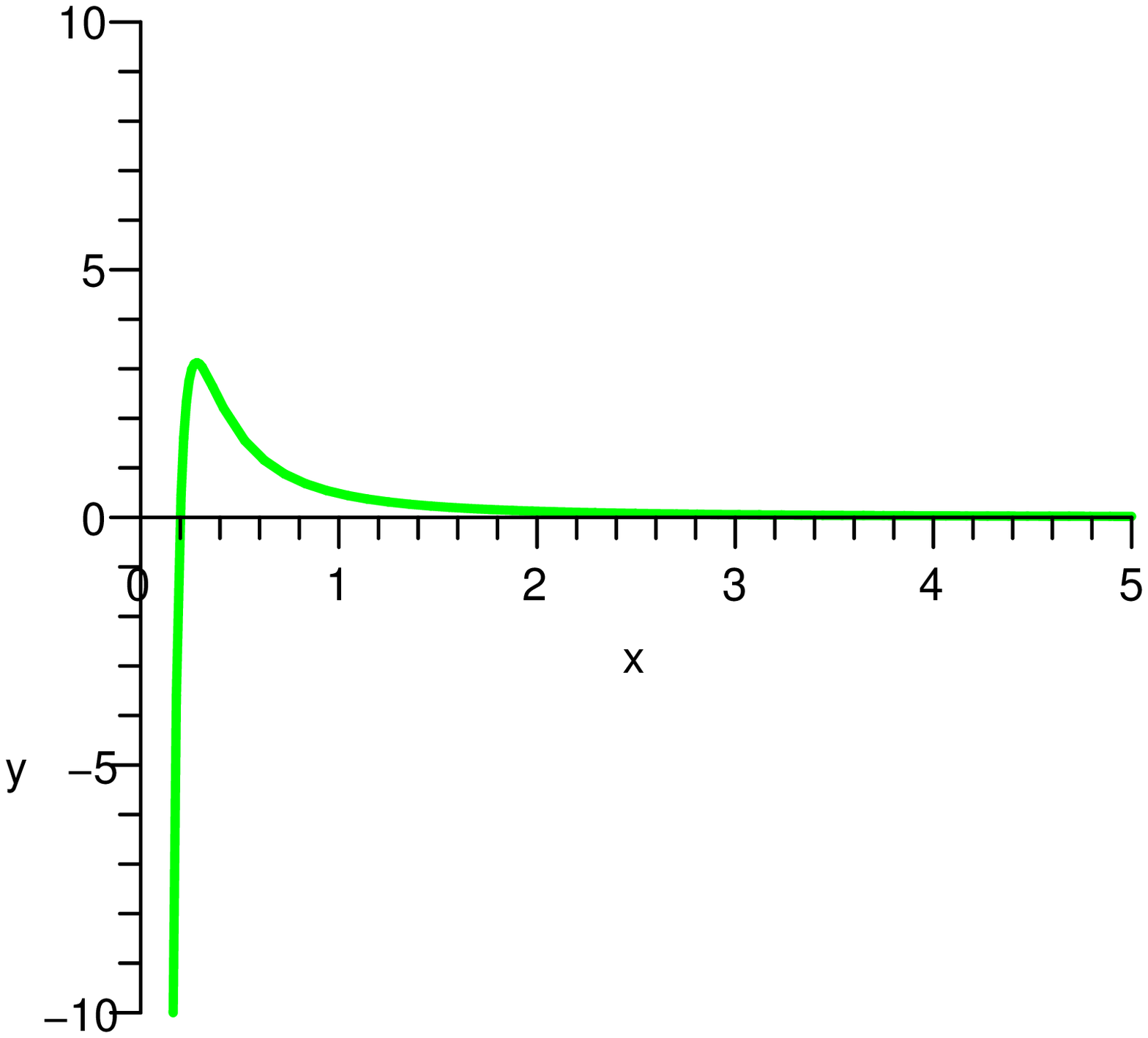}&
\myfigure{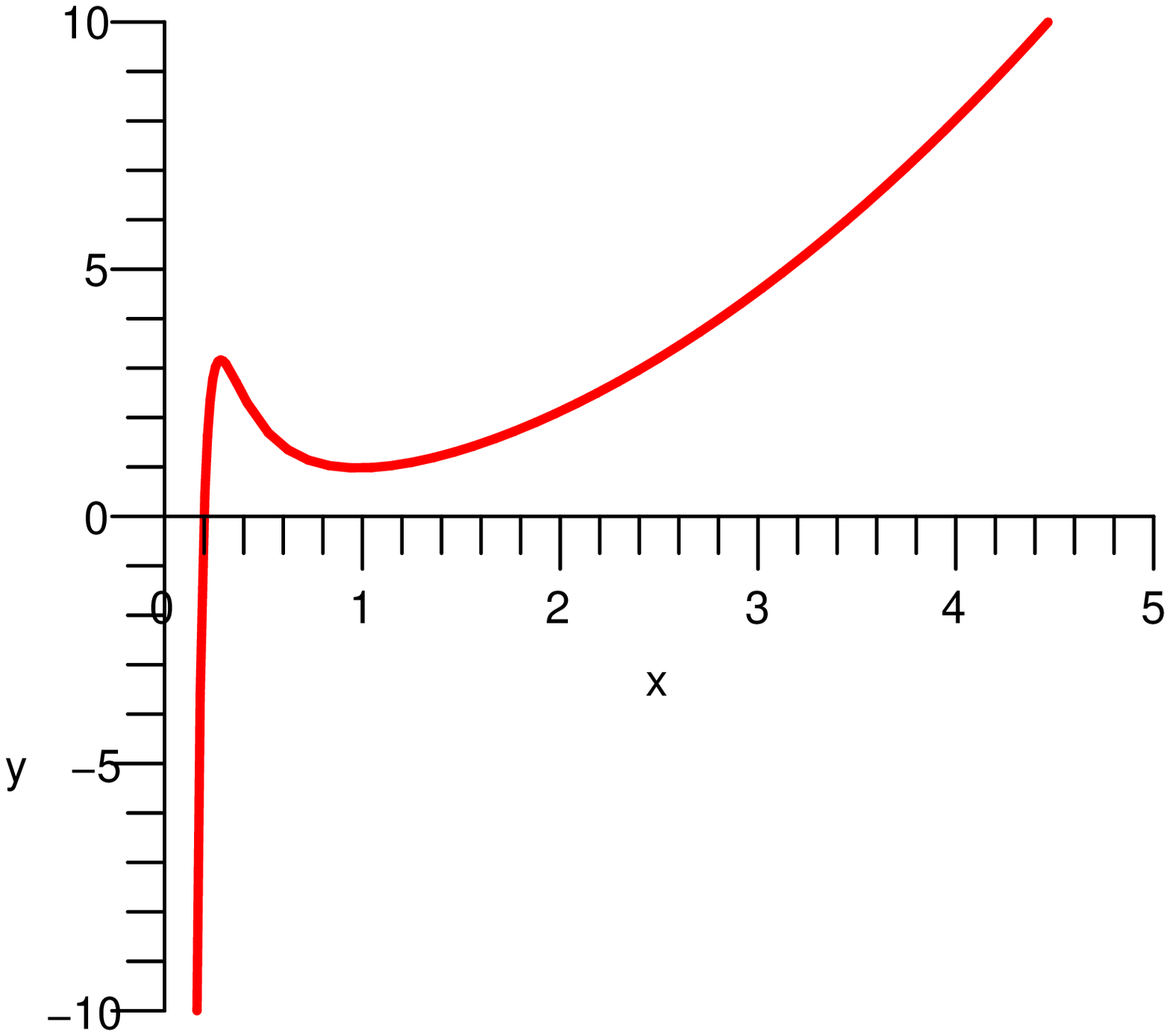}\\
\hline
$\vphantom{\sqrt{\sqrt{\displaystyle\frac12}}}
 \left.a\right)\;\;N=a$&
$\left.b\right)\;\;N=\displaystyle\frac1{a^3},\;\;\varepsilon_0=\frac1{50}$&
$\left.c\right)\;\;N=a+\displaystyle\frac1{a^3},\;\;\varepsilon_0=\frac1{50}$\\
\hline
\end{tabular}
\par\bigskip
{\bf Fig.\thefigure\hspace{2mm}The effective potentials for Eqs.
(\ref{Schr-eq-1}), (\ref{Schr-eq-2}), (\ref{Schr-eq-3}).}
\addtocounter{figure}{1}
\end{center}

2. $Na^3=1$. The gauge is believed to produce the appearance of
$\Lambda$-term in the Einstein equations, since it is the analog of
a more general condition ${\rm det}\|g^{\mu\nu}\|=1$. The equation
of state $p_{(obs)}=-\varepsilon_{(obs)}$. The Schr\"odinger
equation takes the form
\begin{equation}
\label{Schr-eq-2}
-\frac12\frac1{a^4}\frac{d^2\Psi}{da^2}
 +\frac1{a^5}\frac{d\Psi}{d a}
 +\frac1{2a^2}\Psi
 -\frac{\varepsilon_0}{a^4}\Psi=E\Psi.
\end{equation}
Here $\varepsilon_0$ characterizes a contribution of the matter
fields (radiation). If one includes into the model de Sitter false
vacuum with the equation of state $p_{(obs)}=-\varepsilon_{(obs)}$
and the dependence $\varepsilon(a)=\varepsilon_0$, it does not
affect the form of the equation (\ref{Schr-eq-2}) after
redefinition (\ref{E-redef}). Then one could say that vacuum energy
as well as gauge effects are responsible for eigenvalues of $E$.

The effective potential
$U(a)=\displaystyle\frac1{2a^2}-\frac{\varepsilon_0}{a^4}$ depends
on the parameter $\varepsilon_0$. According to modern cosmological
notions, the Universe was created in a metastable state under the
barrier depicted at Fig. 1(b) and then tunneled through the
barrier. The smaller the parameter $\varepsilon_0$ is, the higher
and narrower the barrier becomes. There is a non-zero probability
for arbitrary large values of the scale factor $a$; it means that
the Universe may expand to infinity in spite of the sign ``+'' we
have put before the second term in (\ref{action-1}), which
corresponds to the closed model. It demonstrates that a naive
correspondence between the kind of a cosmological model and the
form of the effective potential has no grounds.

3. $N=a+\displaystyle\frac1{a^3}$. This gauge covers the two
previous cases as asymptotic limits. The equation of state is
\begin{equation}
\label{eq-states-3}
p_{(obs)}=\frac13\,\frac{a^4-3}{a^4+1}\,\varepsilon_{(obs)}.
\end{equation}
At $a\to 0$ the equation gives $p_{(obs)}=-\varepsilon_{(obs)}$; at
$a\to\infty$ it gives
$p_{(obs)}=\displaystyle\frac13\,\varepsilon_{(obs)}$. Again, after
redefinition (\ref{E-redef}) the Schr\"odinger equation looks like
following
\begin{equation}
\label{Schr-eq-3}
-\frac12\left(1+\frac1{a^4}\right)\frac{d^2\Psi}{da^2}
 +\frac1{a^5}\frac{d\Psi}{d a}
 +\frac12a^2\Psi
 +\frac1{2a^2}\Psi
 -\frac{\varepsilon_0}{a^4}\Psi=E\Psi.
\end{equation}
It is easy to check that Eqs. (\ref{Schr-eq-2}), (\ref{Schr-eq-1})
are the asymptotic limits of (\ref{Schr-eq-3}) at $a\to 0$ and
$a\to\infty$ respectively. In this case the Universe is believed to
be filled by some mixture of matter and vacuum. In consequence of
the redefinition (\ref{E-redef}), the value of $E$ is due to matter
contribution as well as gauge effects. Like in a previous case, the
effective potential
$U(a)=\displaystyle\frac12a^2+\frac1{2a^2}-\frac{\varepsilon_0}{a^4}$
depends on the parameter $\varepsilon_0$ and depicted at Fig. 1(c).
The barrier at small $a$ disappears when $\varepsilon_0=0$ and
$\varepsilon_0\ge0.1$. The potential for some value of
$\varepsilon_0$ is shown at Fig. 2. One can see that the potentials
of Eq. (\ref{Schr-eq-2}) (green graph) and of Eq. (\ref{Schr-eq-1})
(blue graph) are asymptotic forms of the potential of Eq.
(\ref{Schr-eq-3}).
\begin{center}
\tabcolsep=1mm
\begin{tabular}{|c|c|c|}
\hline
\rule{0mm}{50mm}
\myfigure{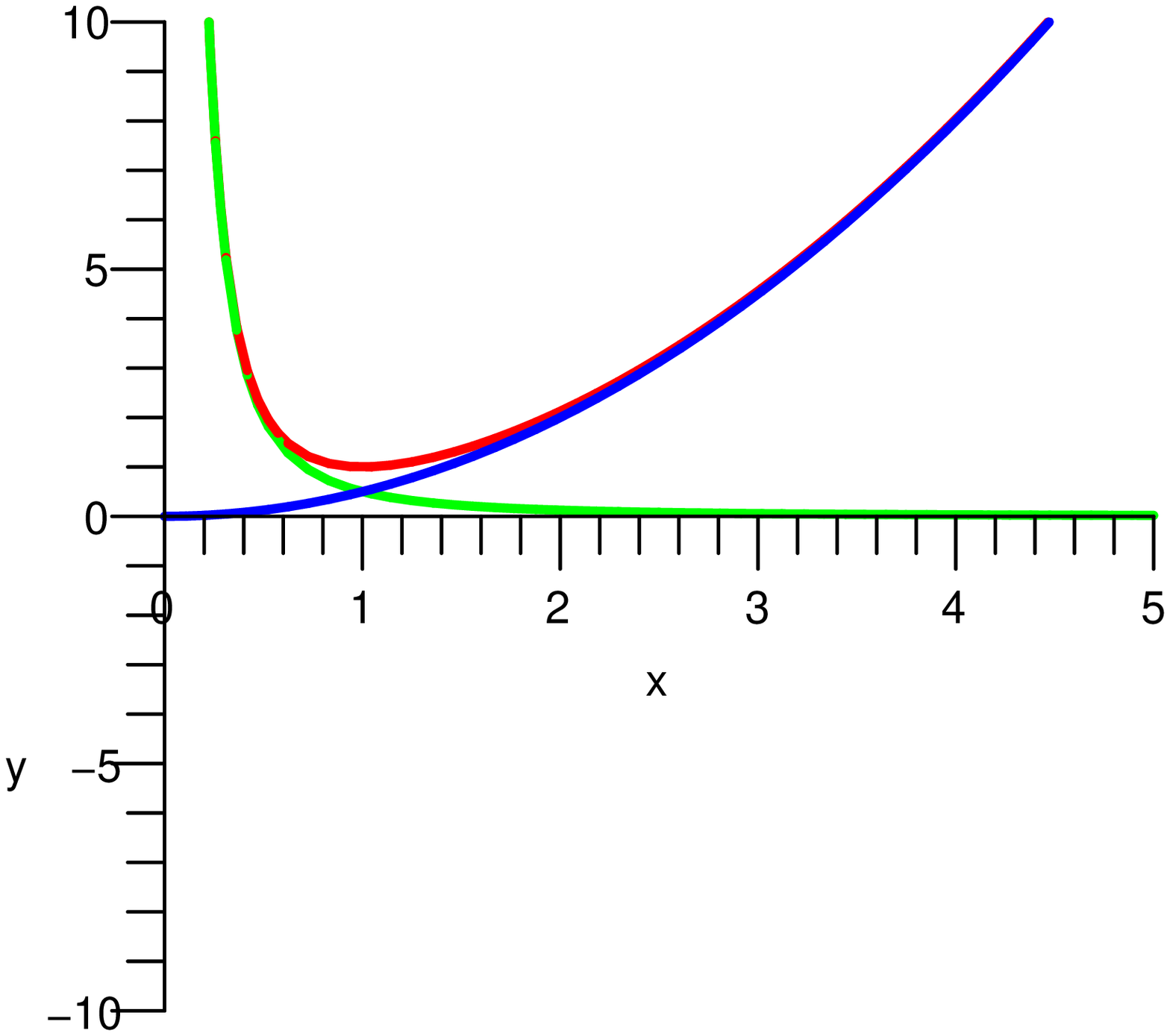}&
\myfigure{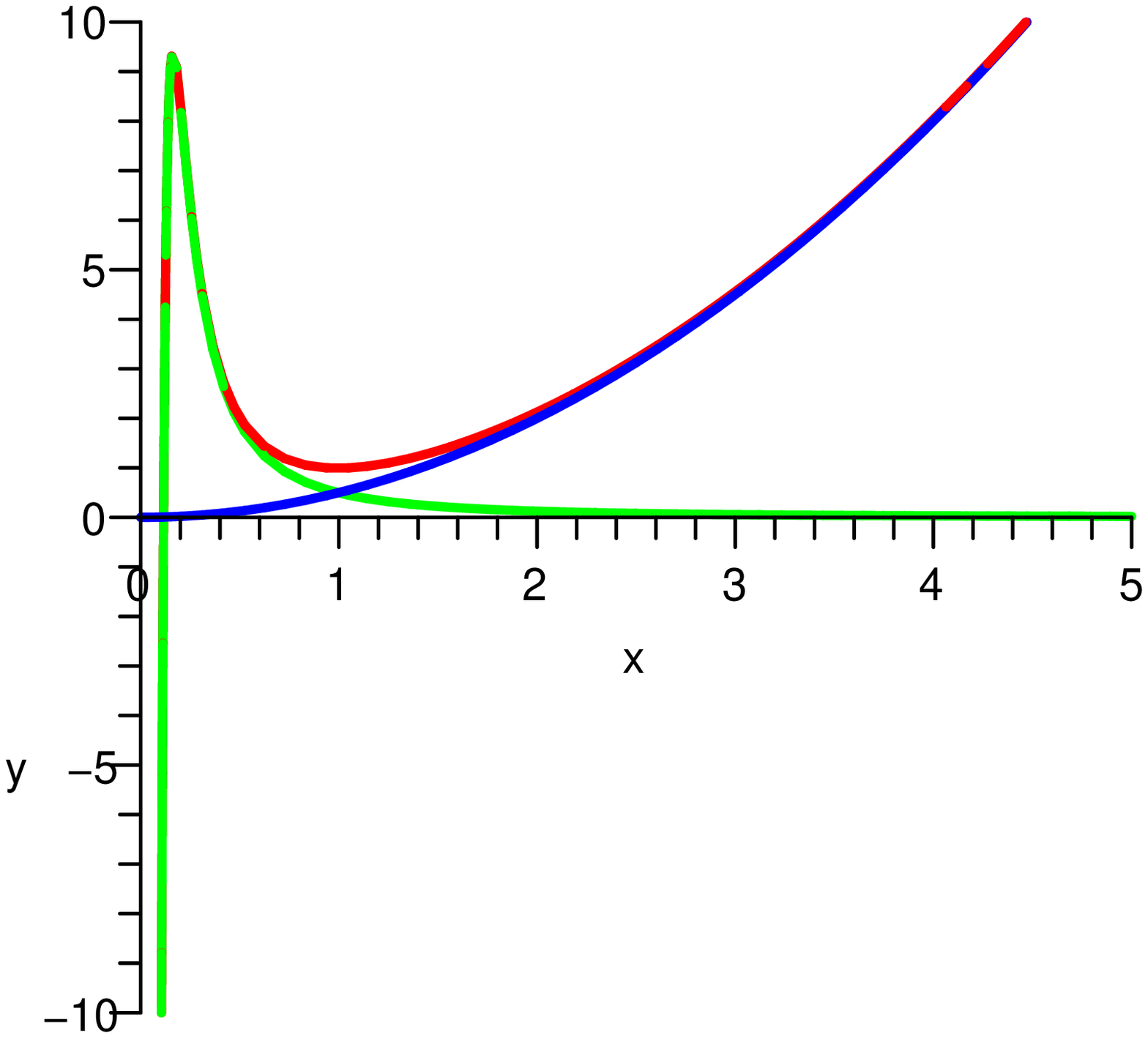}&
\myfigure{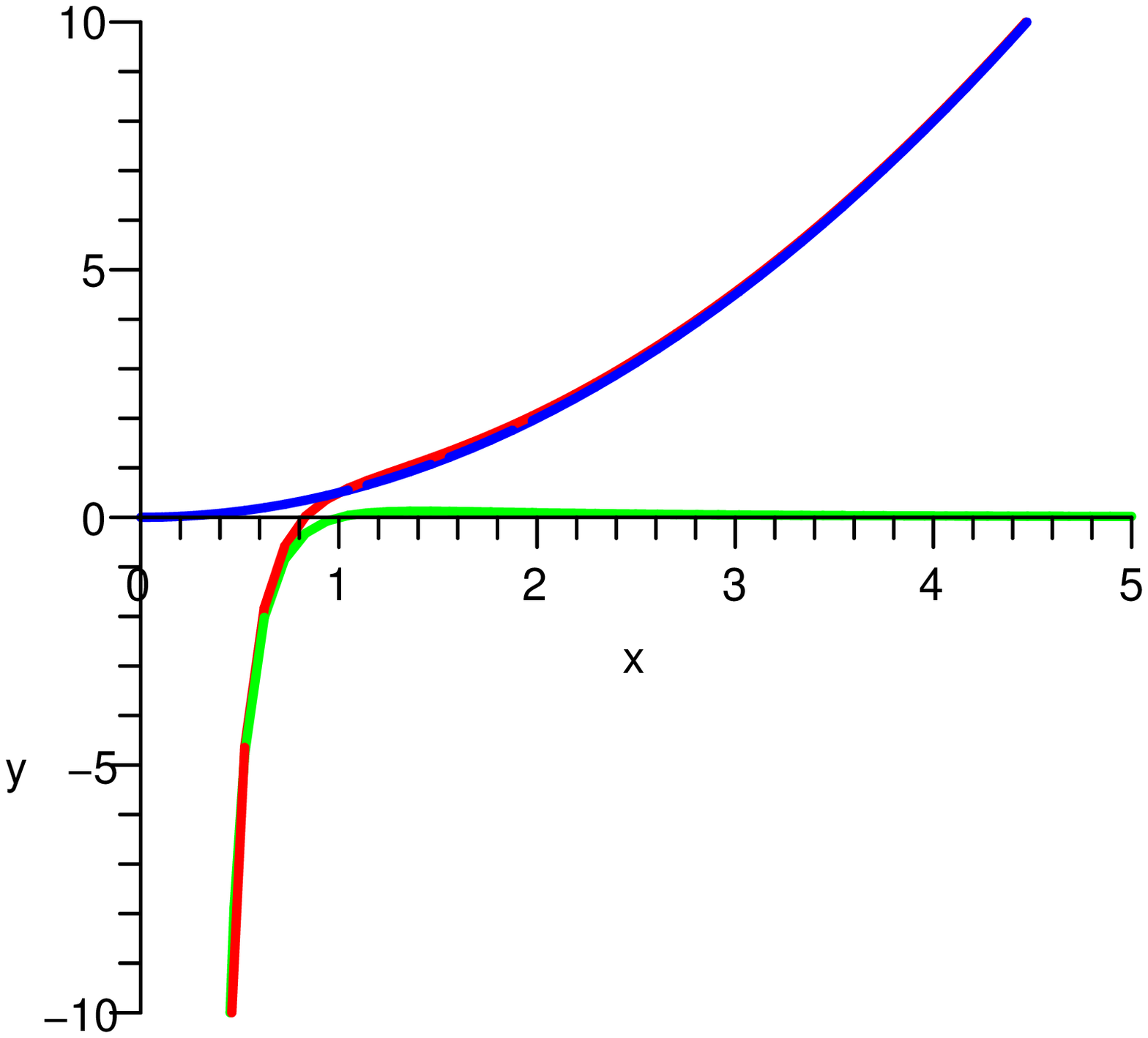}\\
\hline
$\vphantom{\sqrt{\sqrt{\displaystyle\frac12}}}
 \left.a\right)\;\;\varepsilon_0=0$&
$\left.b\right)\;\;\varepsilon_0=\displaystyle\frac1{150}$&
$\left.c\right)\;\;\varepsilon_0=\displaystyle\frac12$\\
\hline
\end{tabular}
\par\bigskip
{\bf Fig.\thefigure\hspace{2mm}The effective potentials for some
values of $\varepsilon_0$.}
\addtocounter{figure}{1}
\end{center}

\section{Numerical solutions}
The Hamiltonian operators in Eqs. (\ref{Schr-eq-1}),
(\ref{Schr-eq-3}) have a discrete spectrum, and one meets no
technical difficulties to obtain numerical solutions to these
equations. The operator in (\ref{Schr-eq-gen}) is Hermitian for an
arbitrary gauge (\ref{gauge}) if the measure in Hilbert space of
solution is taken to be
\begin{equation}
\label{measure}
M(a)=\sqrt{\frac a{f(a)}}.
\end{equation}
One can see that the measure, like the equation itself, is
gauge-dependent.

The standard method of finding eigenvalues and eigenfunctions
consists in the expansion onto a basis functions which are
orthonormal on the interval $[0,\,\infty]$ with the measure
(\ref{measure}):
\begin{equation}
\label{Psi-exp}
\Psi(a)=\sum_n c_n\psi_n^s(a);
\end{equation}
\begin{equation}
\label{eigfun}
\psi_n^s(a)
 =\sqrt{\frac{n!}{(n+s)!}}\,\frac1{\sqrt{M(a)}}\,a^{\frac s2}L_n^s(a)
 =\sqrt{\frac{n!}{(n+s)!}}
  \left(\frac{f(a)}a\right)^{\frac14}a^{\frac s2}L_n^s(a);
\end{equation}
\begin{equation}
\label{ortho}
\int\limits_0^{\infty}\psi_n^{s*}(a)\,\psi_m^s(a)\,M(a)\,d a=\delta_{nm},
\end{equation}
$L_n^s(a)$ are Laguerre polynomials. The problem is reduced to
finding eigenvalues and eigenvectors of the Hamiltonian matrix in
the basis (\ref{eigfun}). The more terms are held in the expansion
(\ref{Psi-exp}), the higher the precision is. The results of
calculations of first five eigenvalues are presented at Table 1.

\begin{center}
{\bf Table 1.}\par
\vspace{3mm}
\tabcolsep=4mm
\renewcommand{\arraystretch}{1.3}
\begin{tabular}{|c|c|c c c c c|}
\hline
\multicolumn{2}{|c|}{Eq. (\ref{Schr-eq-1}), $N=a$}&
 1.5&3.5&5.5&7.50001&9.50008\\
\hline
&$\varepsilon_0=0$&
 2.87886&5.32668&7.66977&9.9591&12.2175\\
\cline{2-7}
&$\varepsilon_0=1/500$&
 2.87846&5.32635&7.66947&9.95882&12.2173\\
\cline{2-7}
&$\varepsilon_0=1/150$&
 2.87754&5.32558&7.66877&9.95817&12.2166\\
\cline{2-7}
&$\varepsilon_0=1/50$&
 2.87489&5.32337&7.66677&9.9563&12.2149\\
\cline{2-7}
Eq. (\ref{Schr-eq-3}),&$\varepsilon_0=1/2$&
 2.77519&5.24152&7.59315&9.88783&12.1496\\
\cline{2-7}
$N=a+\displaystyle\frac1{a^3}$&$\varepsilon_0=1$&
 2.66102&5.15088&7.51266&9.81349&12.0792\\
\cline{2-7}
&$\varepsilon_0=3$&
 2.04887&4.72486&7.14847&9.48369&11.7714\\
\cline{2-7}
&$\varepsilon_0=4$&
 1.59368&4.47069&6.94071&9.29951&11.602\\
\cline{2-7}
&$\varepsilon_0=5$&
 0.972188&4.1924&6.71849&9.1044&11.4236\\
\cline{2-7}
&$\varepsilon_0=7$&
 -1.07592&3.59902&6.25063&8.69468&11.0497\\
\hline
\end{tabular}
\end{center}

One can see that for Eq. (\ref{Schr-eq-1}), $N=a$ the spectrum is
equidistant, the difference between eigenvalues is equal to 2 in
the Plank units (the deviation from this value is entirely due to
calculation inaccuracy).

In the case of Eq. (\ref{Schr-eq-3}),
$N=a+\displaystyle\frac1{a^3}$, the eigenvalues do not differ
significantly for $\varepsilon_0\le\displaystyle\frac1{50}$ and
converge to limiting values at $\varepsilon_0=0$. For
$\varepsilon_0>\displaystyle\frac1{50}$ the spectrum levels tend to
go down into the potential pit. The schematic picture of the
spectrum is shown at Fig.3.

\vspace{15mm}

\begin{center}
\tabcolsep=1mm
\begin{tabular}{|c|c|c|}
\hline
\rule{0mm}{50mm}
\myfigure{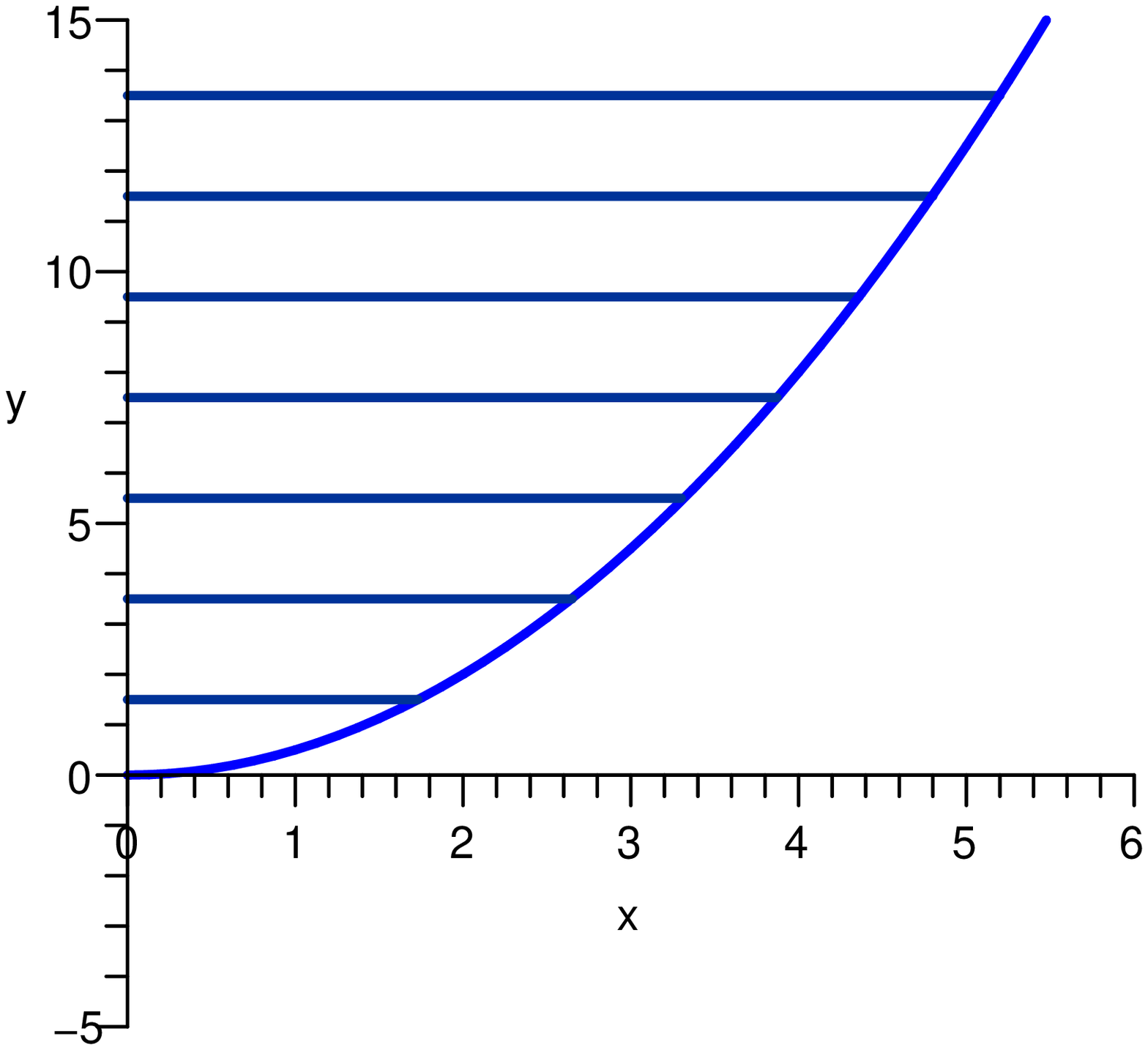}&
\myfigure{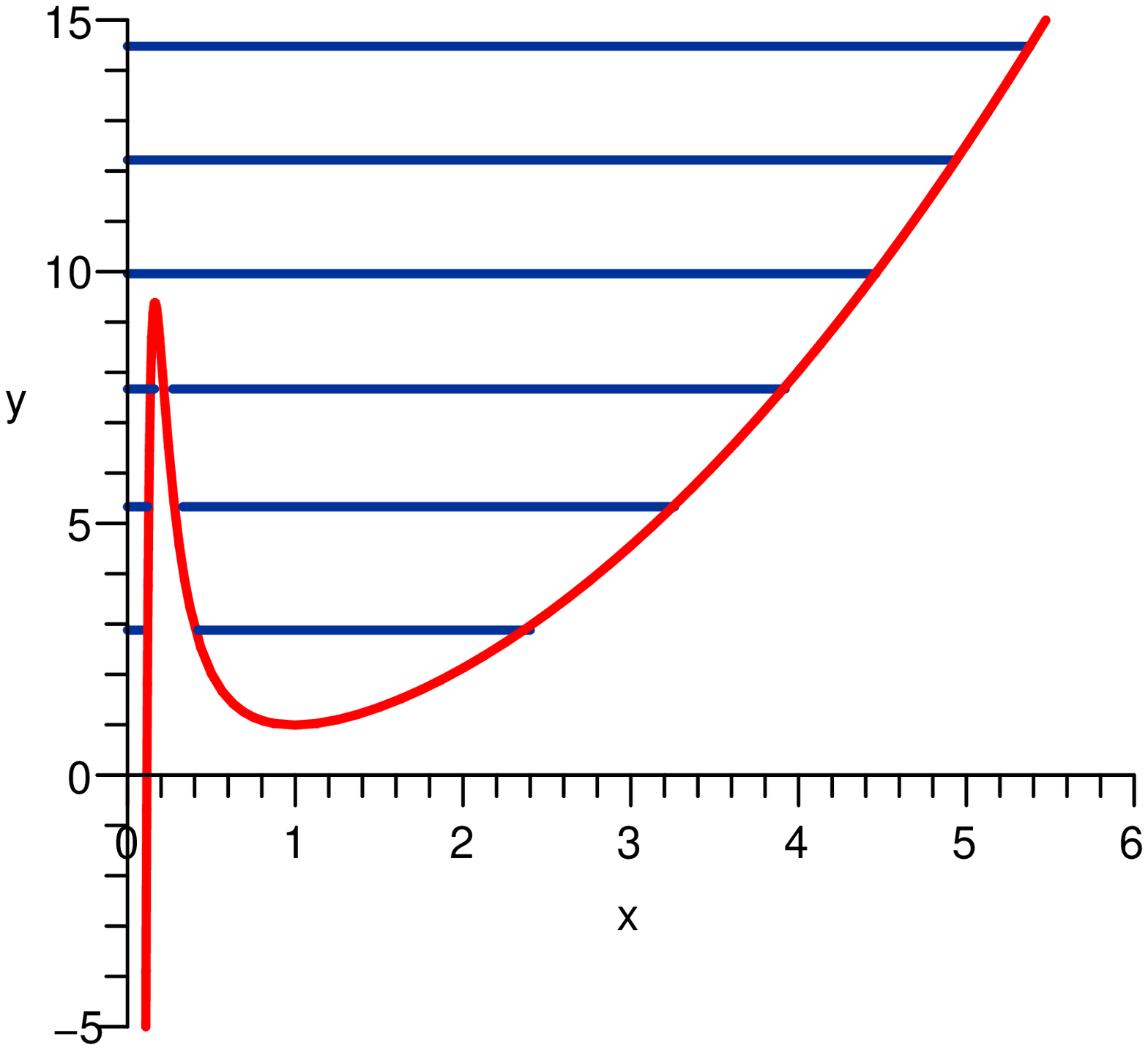}&
\myfigure{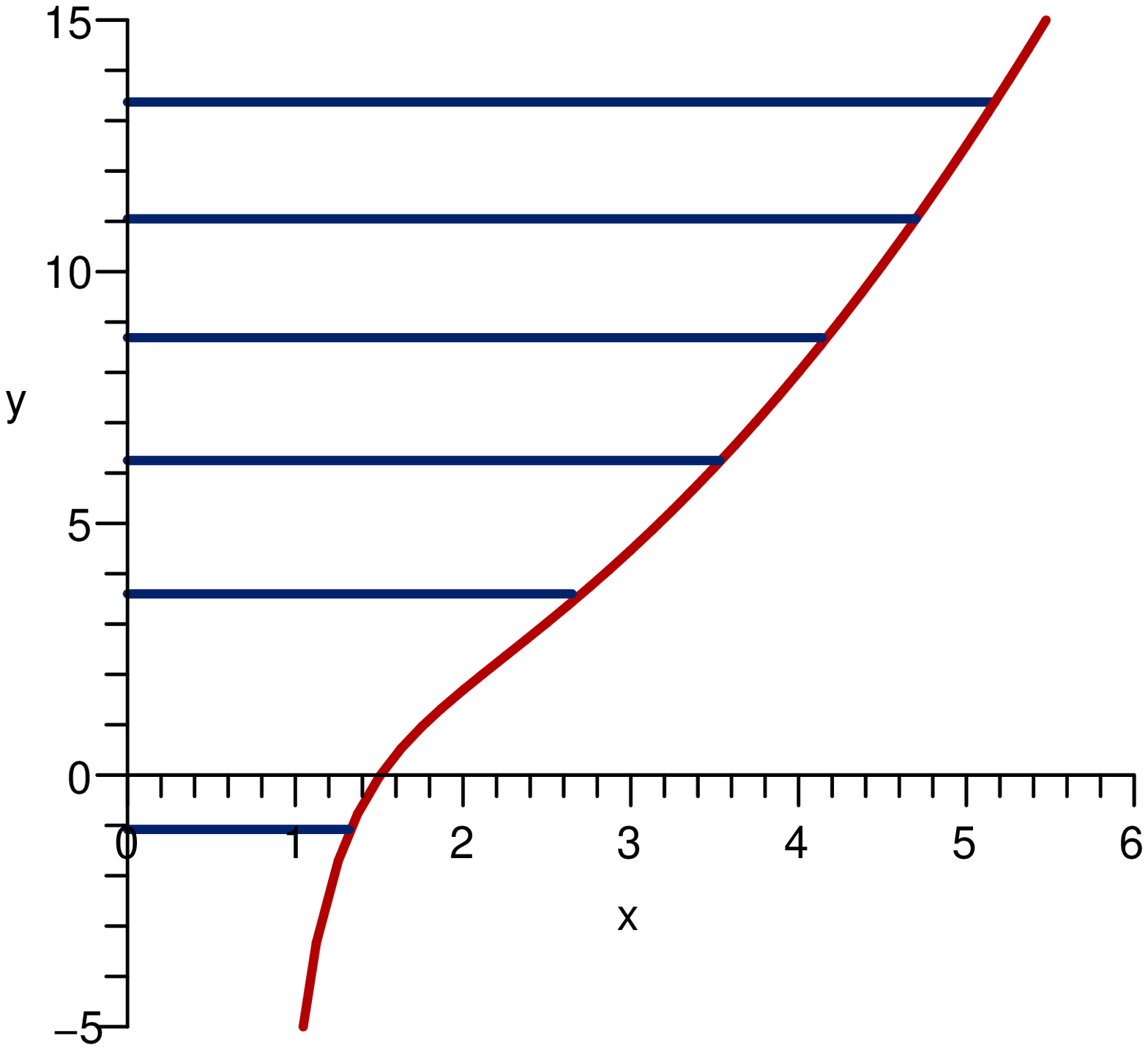}\\
\hline
$\vphantom{\sqrt{\sqrt{\displaystyle\frac12}}}\left.a\right)$
 Eq. (\ref{Schr-eq-1}), $N=a$&
$\left.b\right)$
 Eq. (\ref{Schr-eq-3}), $\varepsilon_0=\displaystyle\frac1{150}$&
$\left.c\right)$
 Eq. (\ref{Schr-eq-3}), $\varepsilon_0=0$\\
\hline
\end{tabular}
\par\bigskip
{\bf Fig.\thefigure\hspace{2mm}The spectrum levels for some potentials.}
\addtocounter{figure}{1}
\end{center}

\vspace{15mm}

Fig. 4--6 pictures the probability distributions for the first
(ground state), third and fifth solutions to Eq. (\ref{Schr-eq-1})
and Eq. (\ref{Schr-eq-3}) when
$\varepsilon_0=\displaystyle\frac1{150}$ and $\varepsilon_0=7$. One
can see that at the qualitative level the probability distributions
do not significantly differ. The peak of the probability
distribution in the all cases tends to shift to large values of the
scale factor $a$ for larger eigenvalues of $E$. One could expect
this result since the matter and gauge effects contribute to the
value of $E$. So, when the energy of matter increases, there may be
enough probability for the scale factor to reach large values.

\newpage
\begin{center}
\tabcolsep=1mm
\begin{tabular}{|c|c|c|}
\hline
\rule{0mm}{50mm}
\myfigure{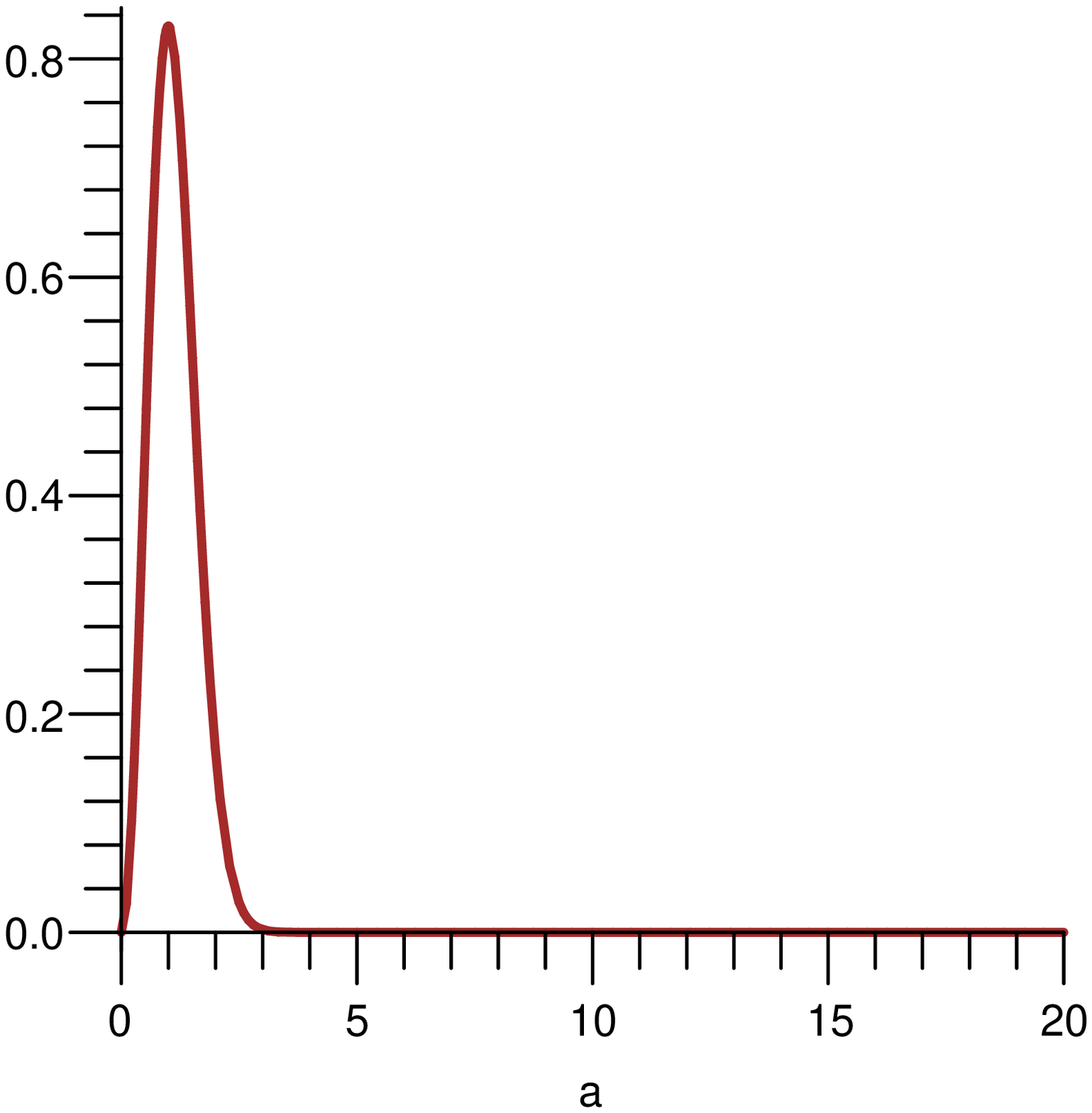}&
\myfigure{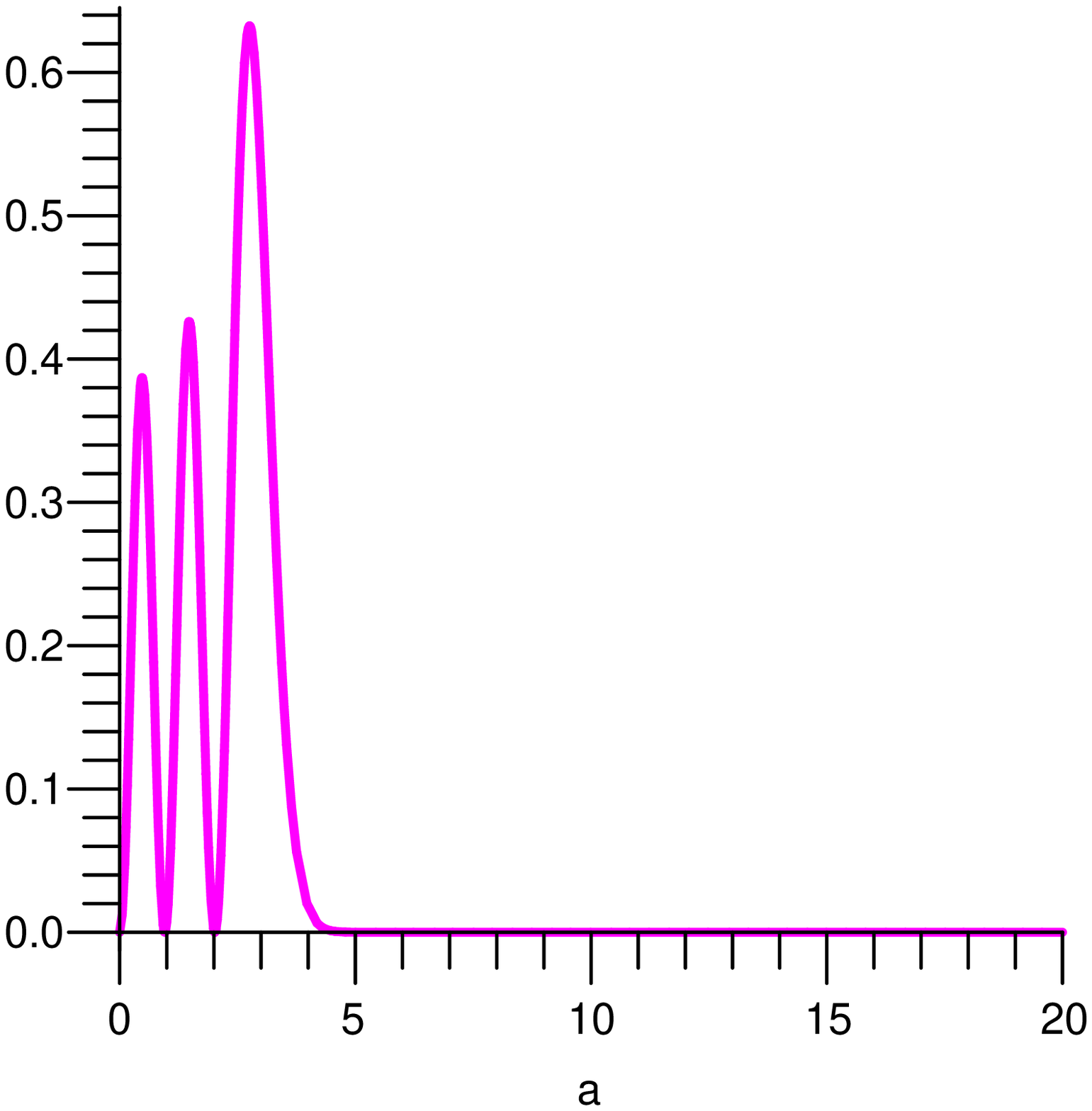}&
\myfigure{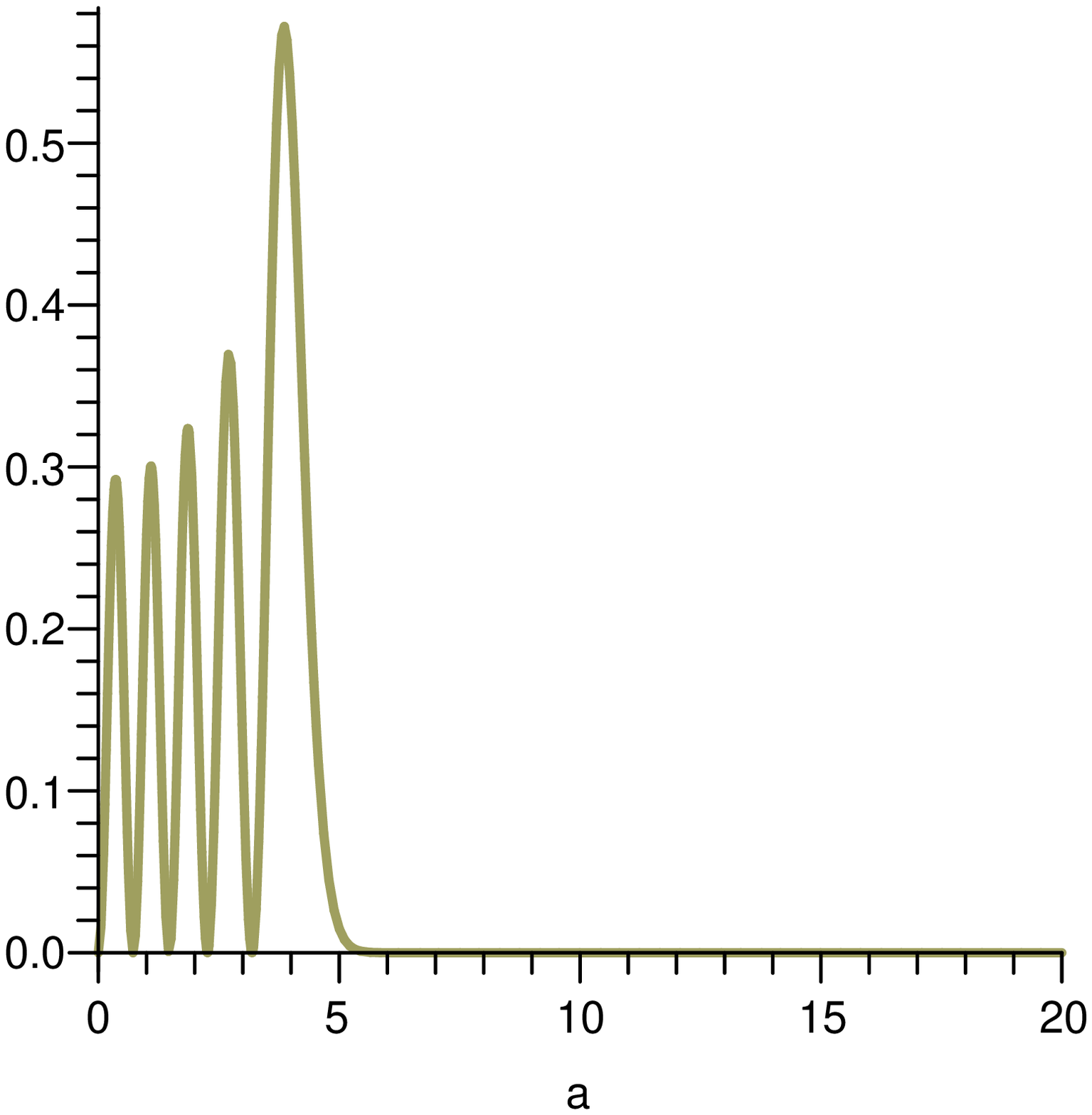}\\
\hline
$\vphantom{\displaystyle\frac12}
 \left.a\right)\;\;\left|\Psi_0\right|^2,\;\;E_0=1.5$&
$\left.b\right)\;\;\left|\Psi_2\right|^2,\;\;E_2=5.5$&
$\left.c\right)\;\;\left|\Psi_4\right|^2,\;\;E_4=9.50008$\\
\hline
\end{tabular}
\par\bigskip
{\bf Fig.\thefigure\hspace{2mm}The probability distributions for
solutions to Eq. (\ref{Schr-eq-1}), $N=a$}
\addtocounter{figure}{1}
\end{center}
\vspace{2mm}
\begin{center}
\tabcolsep=1mm
\begin{tabular}{|c|c|c|}
\hline
\rule{0mm}{50mm}
\myfigure{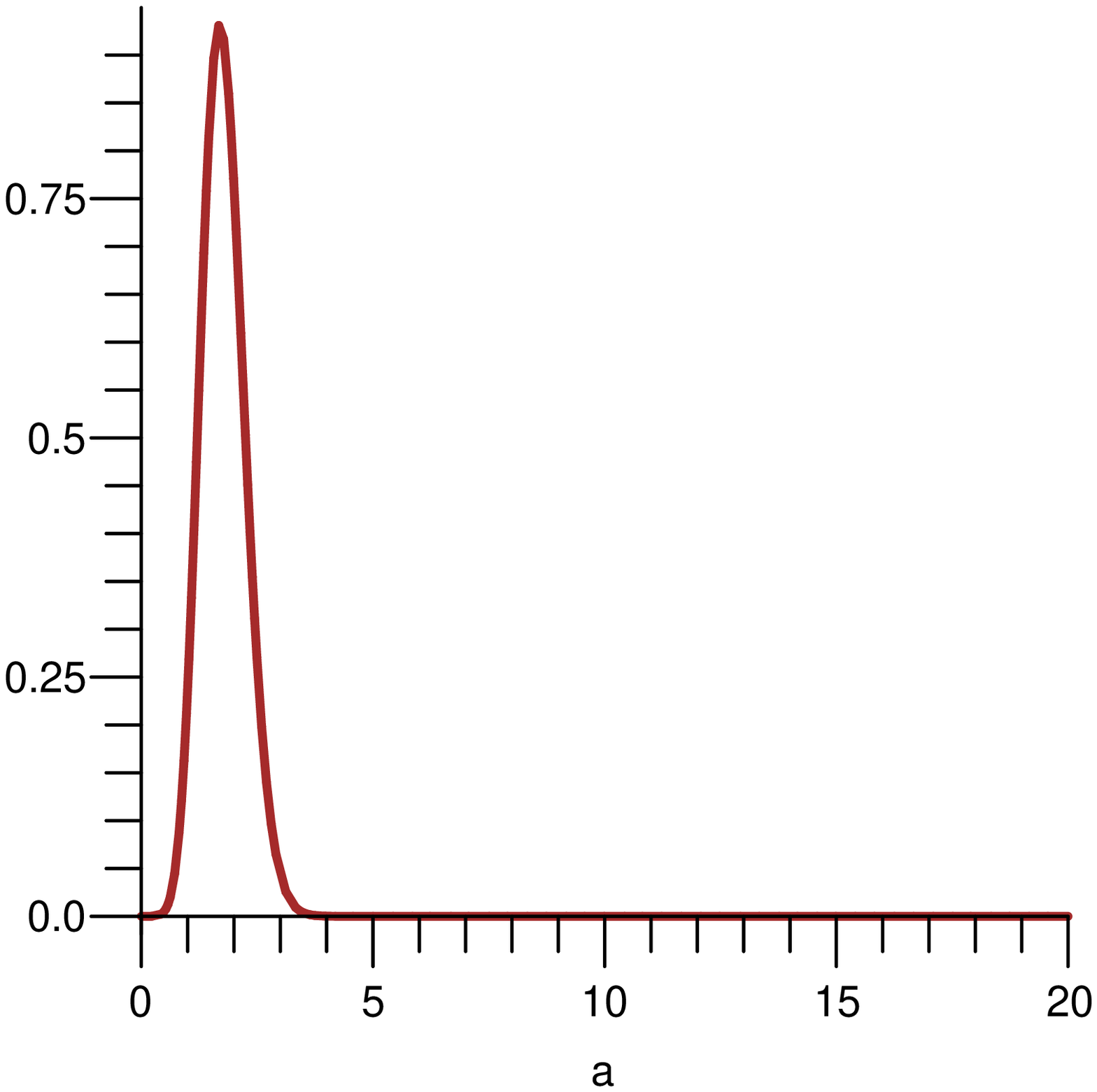}&
\myfigure{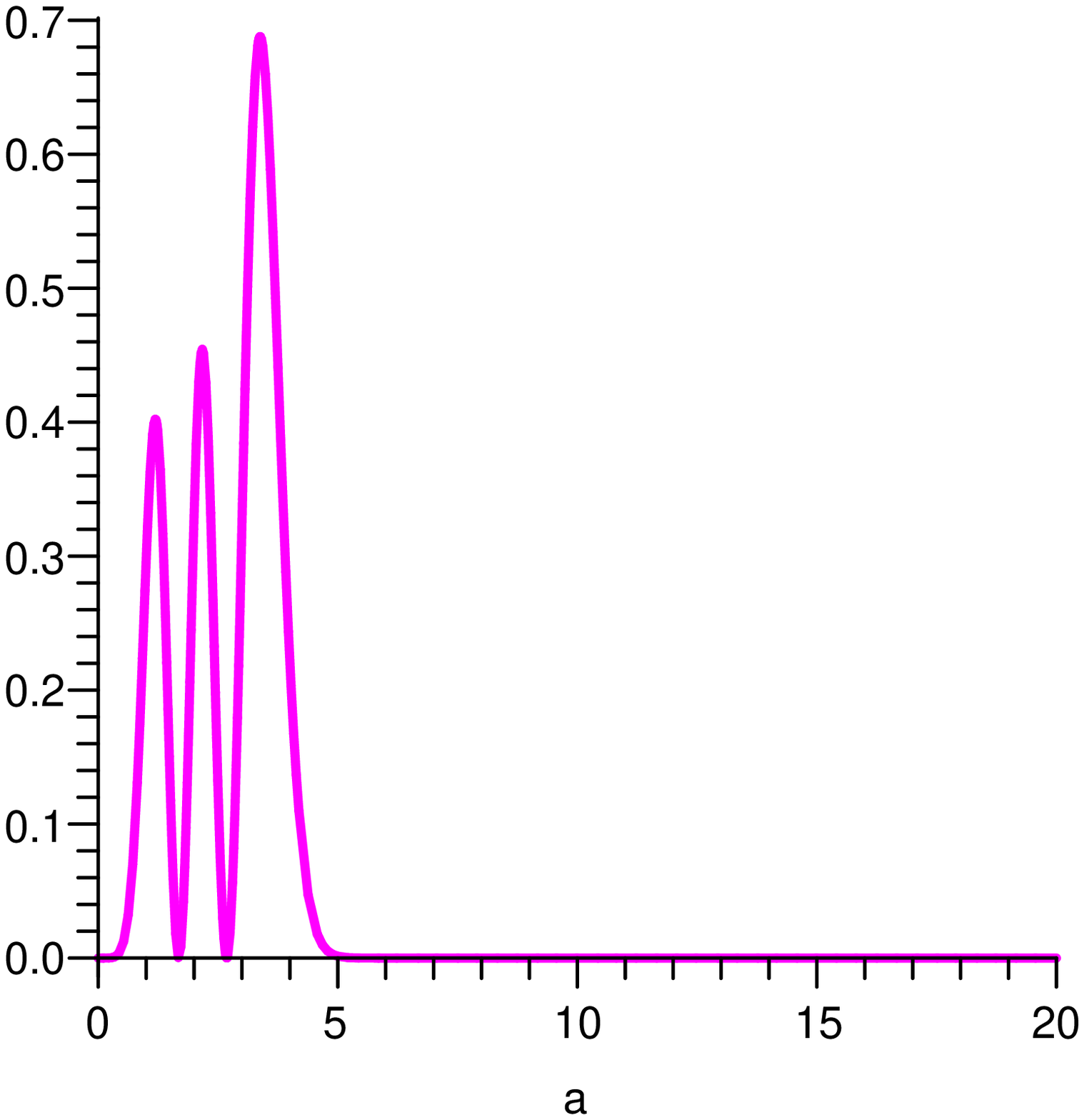}&
\myfigure{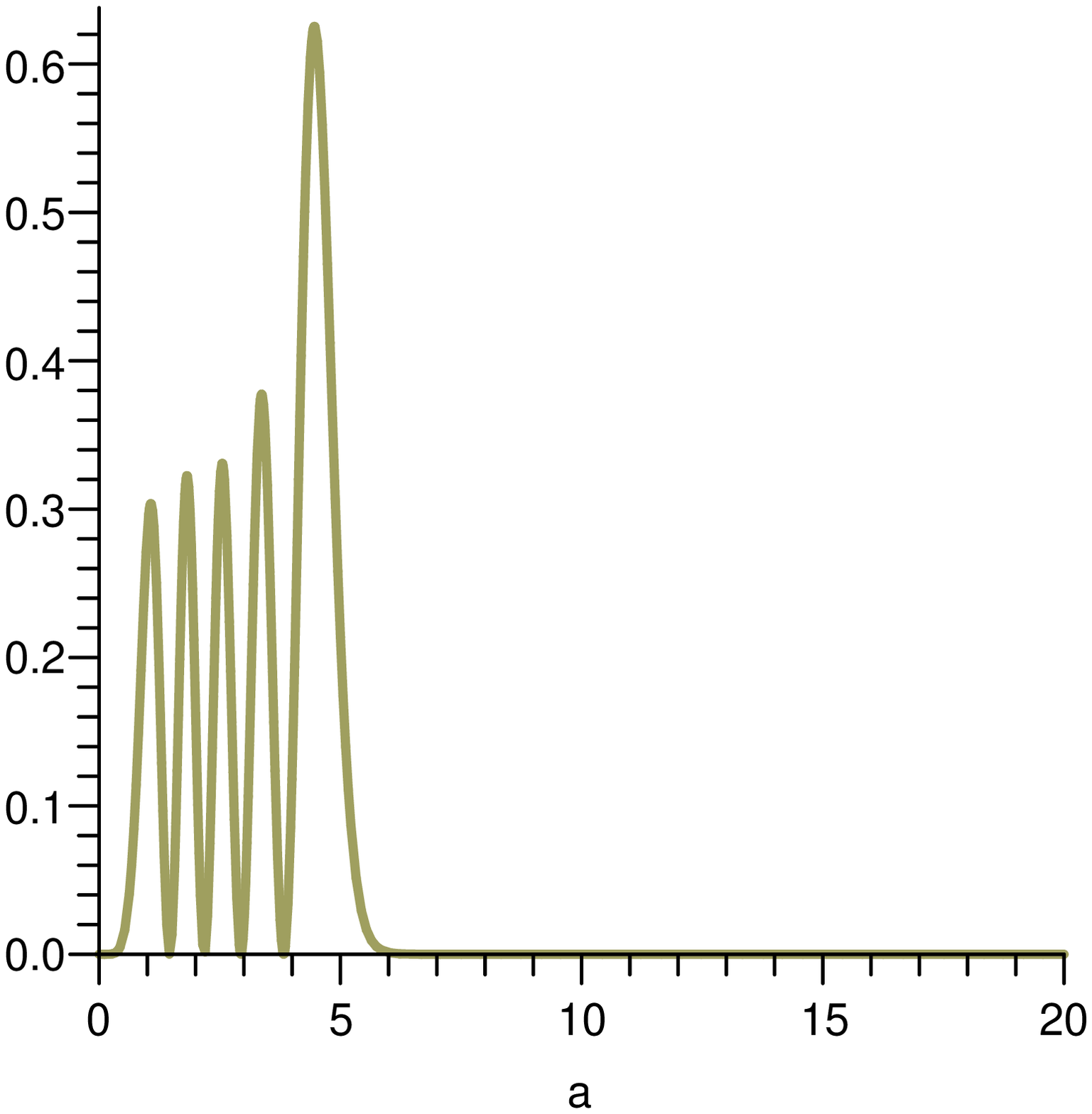}\\
\hline
$\vphantom{\displaystyle\frac12}
 \left.a\right)\;\;\left|\Psi_0\right|^2,\;\;E_0=2.87754$&
$\left.b\right)\;\;\left|\Psi_2\right|^2,\;\;E_2=7.66877$&
$\left.c\right)\;\;\left|\Psi_4\right|^2,\;\;E_4=12.2166$\\
\hline
\end{tabular}
\par\bigskip
{\bf Fig.\thefigure\hspace{2mm}The probability distributions for
solutions to Eq.(\ref{Schr-eq-3}), $N=a+\displaystyle\frac1{a^3}$,
$\varepsilon_0=\displaystyle\frac1{150}$}
\addtocounter{figure}{1}
\end{center}
\vspace{2mm}
\begin{center}
\tabcolsep=1mm
\begin{tabular}{|c|c|c|}
\hline
\rule{0mm}{50mm}
\myfigure{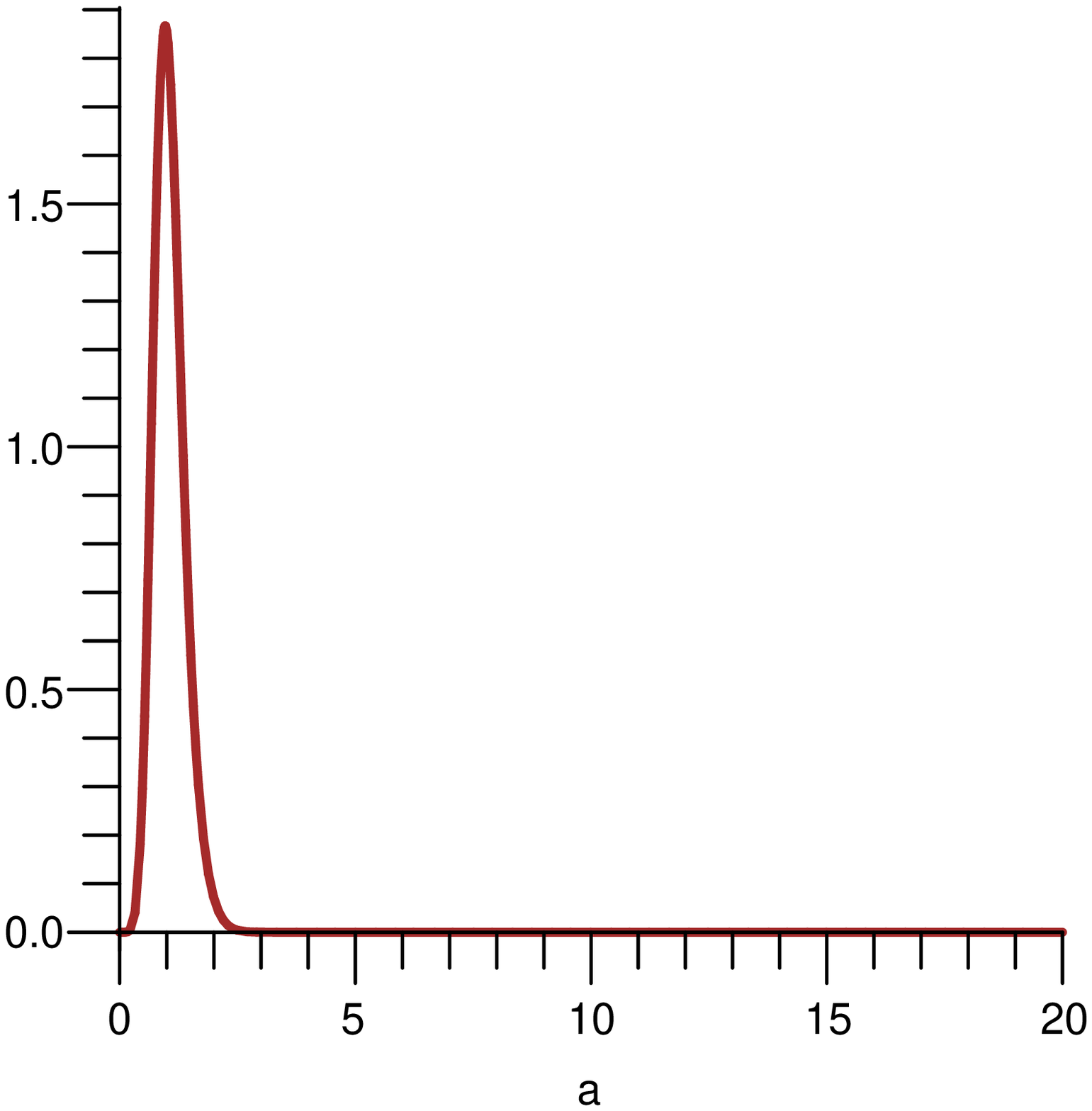}&
\myfigure{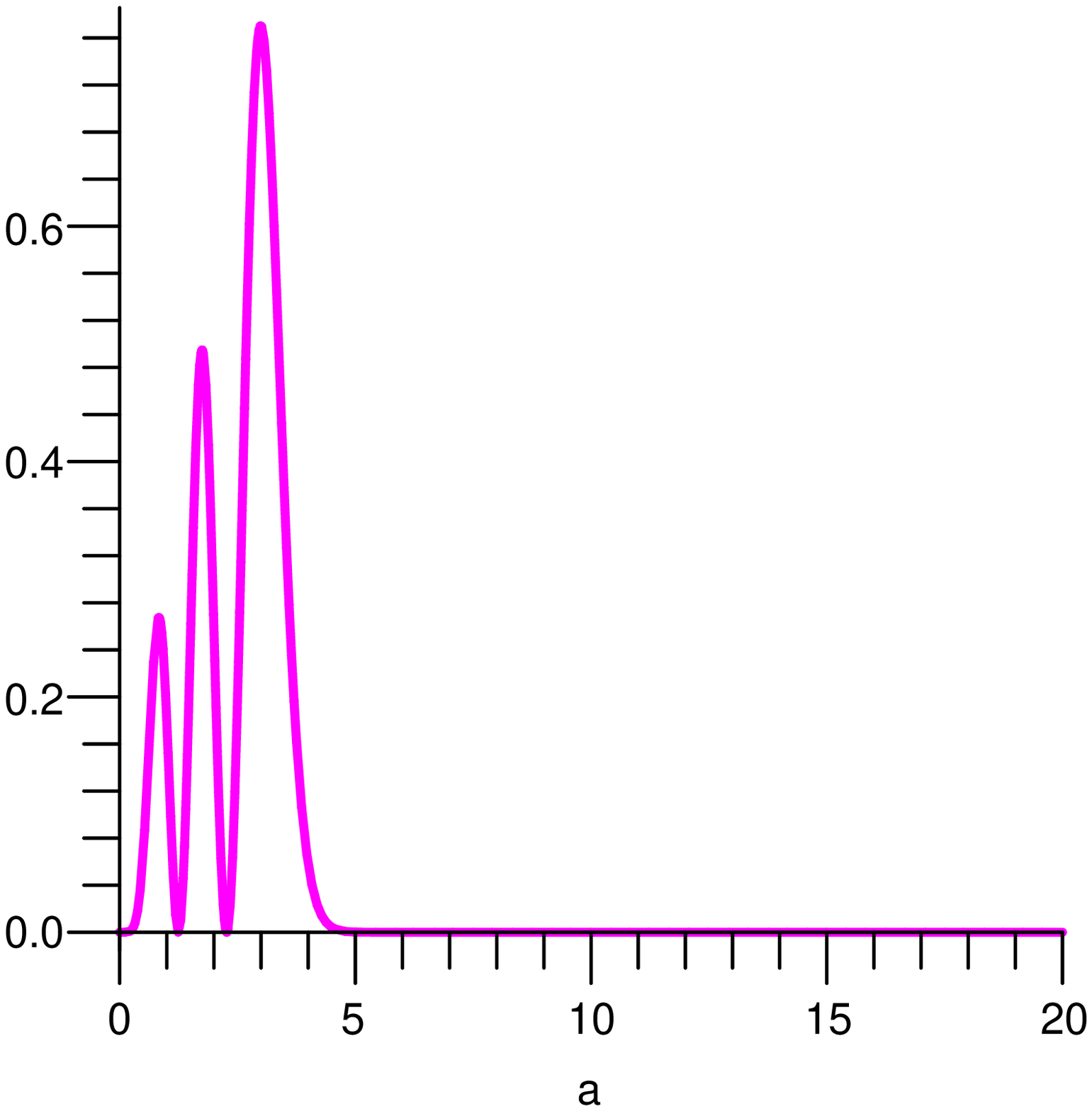}&
\myfigure{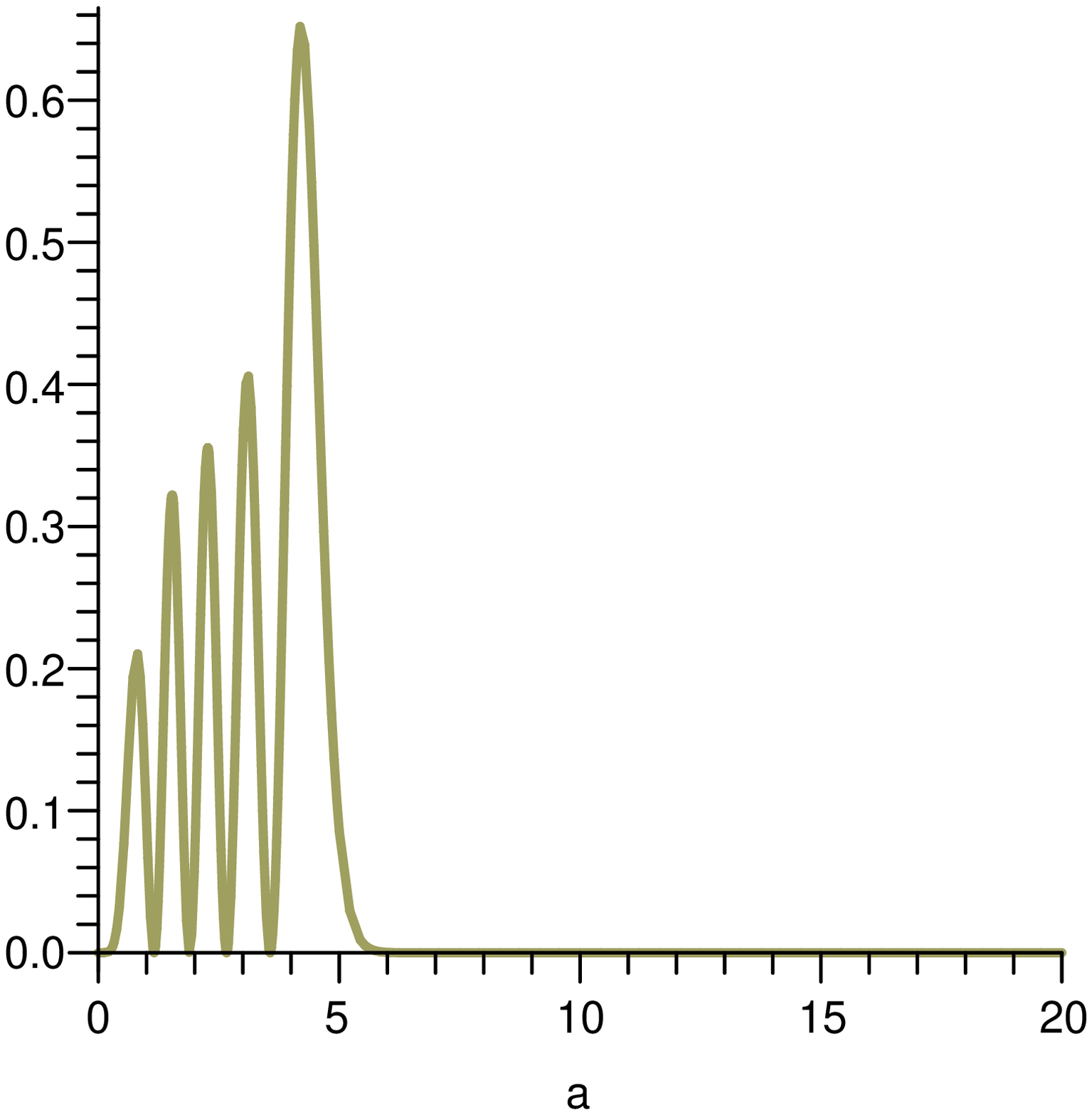}\\
\hline
$\vphantom{\displaystyle\frac12}
 \left.a\right)\;\;\left|\Psi_0\right|^2,\;\;E_0=-1.07592$&
$\left.b\right)\;\;\left|\Psi_2\right|^2,\;\;E_2=6.25063$&
$\left.c\right)\;\;\left|\Psi_4\right|^2,\;\;E_4=11.0497$\\
\hline
\end{tabular}
\par\bigskip
{\bf Fig.\thefigure\hspace{2mm}The probability distributions for
solutions to Eq. (\ref{Schr-eq-3}), $N=a+\displaystyle\frac1{a^3}$,
$\varepsilon_0=7$}
\addtocounter{figure}{1}
\end{center}

\newpage
\section{Concluding remarks}
We should recognize that we have considered a very simple model and
the obtained results are not of high degree of generality. The
present work is just a small step ``to find the way''.

We have seen that the second gauge condition,
$N=\displaystyle\frac1{a^3}$, leads to a continuous spectrum of
eigenvalues of the Schr\"odinger equation (\ref{Schr-eq-gen}),
while the two other gauges, $N=a$ and
$N=a+\displaystyle\frac1{a^3}$, leads to a discrete spectrum, in
other words, the second case is substantially different. It seems
that one should seek for the reason in the structure of spacetime.
Indeed, the gauge $N=\displaystyle\frac1{a^3}$ corresponds to the
Universe in which the interval of proper time between two
subsequent spacelike hypersurfaces tends to zero as $a\to\infty$,
meantime it is not the case for the two other gauges. Since any
gauge condition determines the form of the effective potential,
this circumstance require a more careful exploration. It would be
interesting to study the gauge $N=1+\displaystyle\frac1{a^3}$, for
which at $a\to\infty$ the reference frame becomes a synchronous one
($N=1$) and the equation of state of the observer subsystem at
$a\to\infty$ is that of dust: $p_{(obs)}=0$.

The resemblance of probability distributions for solutions to Eqs.
(\ref{Schr-eq-1}), (\ref{Schr-eq-3}) also deserves our attention.
It demonstrates that one can reveal some relation among solutions
for certain classes of gauge conditions. Let us note that this
problem is well-known in the Wheeler -- DeWitt quantum
geometrodynamics, and the question how solutions to the Wheeler --
DeWitt equation are related, was discussed as soon as
parametrization noninvariance of this theory had been realized.
Then Halliwell \cite{Hall} proposed to restrict the class of
admissible parametrizations. Since parametrization and gauge
conditions have a unified interpretation \cite{SS2}, it implies
also a restriction of the class of admissible gauge conditions,
i.e. such an approach implies that it is permissible to describe
the Universe in only one or several ``privileged'' reference
frames. This way seems to be artificial since we do not know for
sure what reference frame is privileged. Our point of view is that
we face a new problem of finding classes of gauge conditions within
which solutions to the Schr\"odinger equation are stable enough
with respect to a choice of gauges, the determination of the
classes seems to be inseparable from our understanding of spacetime
structure.

\end{document}